\documentclass[aps,prb,preprint,groupedaddress]{revtex4}

\usepackage{graphicx}

\begin{document}

\title{Theory of electrical rectification in a molecular monolayer}

\author{C. Krzeminski, C. Delerue$^{*}$, G. Allan, D. Vuillaume}

\affiliation{Institut d'Electronique et de Micro\'electronique du Nord, D\'epartement Institut Sup\'erieur d'Electronique du Nord, 41 Boulevard Vauban, 59046 Lille C\'edex, France}

\author{R. M. Metzger}

\affiliation{Laboratory for Molecular Electronics, Department of Chemistry, University of Alabama, P.O. Box 870336, Tuscaloosa, Alabama 35487-0336}

\email[*]{Christophe.Delerue@isen.fr}

\vspace{10cm}

\begin{abstract}
The current-voltage characteristics in Langmuir-Blodgett monolayers of $\gamma$-hexadecylquinolinium tricyanoquinodimethanide (C$_{16}$H$_{33}$Q-3CNQ) sandwiched between Al or Au electrodes is calculated, combining ab initio and self-consistent tight binding techniques. The rectification current depends on the position of the LUMO and HOMO relative to the Fermi levels of the electrodes as in the Aviram-Ratner mechanism, but also on the profile of the electrostatic potential which is extremely sensitive to where the electroactive part of the molecule lies in the monolayer. This second effect can produce rectification in the direction opposite to the Aviram-Ratner prediction.
\end{abstract}


\maketitle

\section{Introduction}

Molecular electronics has gained large-scale interest in recent years. A promising system is the electrical rectifier based on suitably engineered molecules. At the origin of this idea, Aviram and Ratner \cite{Aviram74} proposed in 1974 to use D-$\sigma$-A molecules, where D and A are, respectively, an electron donor and an electron acceptor, and $\sigma$  is a covalent "sigma" bridge. Electrical rectification was observed only recently, in particular in Langmuir-Blodgett (LB) multilayers or monolayers of $\gamma$-hexadecylquinolinium tricyanoquinodimethanide (C$_{16}$H$_{33}$Q-3CNQ, {\bf 1}, Fig.  \ref{fig:fig1}) sandwiched between metallic electrodes \cite{Ashwell93,Metzger97,Vuillaume99,Metzger99,Metzger01}. Even if these results represent an important progress to achieve molecular electronics, the physical mechanism responsible for the rectification is not clear. One critical issue is to know if the Aviram-Ratner model can be applied to {\bf 1}, because it is a D(+)-$\pi$-A(-) molecule \cite{Metzger99}. On the theoretical side, these molecular diodes are complex systems, characterized by large and inhomogeneous electric fields, which result from the molecular dipoles \cite{Metzger99} in the layer, the applied bias, and the screening induced by the molecules themselves and the metallic electrodes. A theoretical treatment of these effects is presently lacking, as it requires a self-consistent resolution of the quantum-mechanical problem, including the effect of the applied bias on the electronic structure. Our aim in this paper is to present such a theory, and its application to the systems experimentally studied in Refs. \cite{Metzger97,Vuillaume99,Metzger01}. Combining ab initio and semi-empirical calculations, we show that the direction of easy current flow (rectification current) depends not only on the position of the HOMO and LUMO, relative to the Fermi levels of the metal electrodes before bias is applied, but also on their shift after the bias is applied: this situation is more complex than the Aviram-Ratner mechanism, and can provide a rectification current in the opposite direction. We calculate that the electrical rectification results from the asymmetric profile of the electrostatic potential across the system. We obtain that the C$_{16}$H$_{33}$ tail plays an important role in this asymmetry, and we predict a more symmetric $I(V)$ curve in the case of molecules with a small alkyl chain. Quite generally, our work emphasizes the importance of the electrostatic potential profile in a molecular system and suggests that this profile could be chemically designed to build new devices.

\section{ELECTRONIC STRUCTURE AND POTENTIAL}
To calculate the electronic structure and the geometry of the free molecules (gas phase), we have performed ab initio calculations, using the DMOL code \cite{Cerius2} in the local density approximation (LDA) with the functional of Ref. \cite{Vosko80}, and in the generalized gradient approximation (GGA) of Ref. \cite{Perdew92}. The two approximations give very close results. We use a double numerical basis set(two atomic orbitals for each occupied orbital in the free atom), together with polarization functions ($2p$ for H, and $3$d for N and C). As the molecular diode is a complex system, which cannot be treated in LDA, we have developed self-consistent tight binding (TB) calculations, as described in Ref. \cite{Krzeminski99}. We start from the GGA optimized geometries. C and N atoms are represented by one s and three p atomic orbitals, and H atoms by one s orbital. The diagonal terms of the Hamiltonian matrix are charge-dependent \cite{Krzeminski99,Lannoo74} (in atomic units):

\begin{equation}
H_{i \alpha,i \alpha}=H^{0}_{i \alpha,i \alpha}-\sum_{j}\frac{Q_{j}}{\sqrt{R_{ij}^{2}+(\frac{e^{2}}{U_{0}^{2}})}}-eV_{ext}(i)
\label{eq:eq_one}
\end{equation}

where $\alpha$ is the orbital index, $i$ and $j$ are atomic indices ($j$ runs over all the atoms of the system), $Q_{j}$ is the net charge on atom $j$, $R_{ij}$ is the distance between atoms $i$ and $j$, $U_{0}$ is the intra-atomic Coulomb energy and the quantities $H^{0}_{i \alpha,i \alpha}$  define the s and p atomic levels. $V_{ext}(i)$ is the potential induced by the charges on the electrodes. It is calculated using a self-consistent solution of the Schr\"{o}dinger and Poisson equations, with the boundary condition imposed by the applied bias. The second term in eqn. \ref{eq:eq_one} is evaluated using the method of Ewald \cite{Ewald21}, taking into account the two-dimensional periodicity of the LB film. The molecules are ordered on a 9\AA $\times$ 6.5\AA  \quad  rectangular lattice, corresponding to a molecular area of 58.5 \AA$^{2}$, close to the experimental value of $\sim$ 50 \AA$^{2}$ \quad  \cite{Metzger97}. The molecules are inclined by 45$^{\circ}$, as measured experimentally \cite{Metzger97}. Chemical interactions between molecules are supposed to be negligible. The LB film is sandwiched between two planar Al or Au electrodes, which we put at 3.2 \AA  \quad from the edges of the molecule, a reasonable value for molecules in weak interaction with a metallic surface \cite{Koch99}. We applied our calculations to {\bf 1}, but {\bf 3} also to molecules with a smaller alkyl chain or without alkyl chain (Q-3CNQ, {\bf 2}) for reasons which will become clear later. As expected, the electronic structure cose to the gap does not depend on the length of the alkyl chain for the free molecules. We predict a large dipole moment of 21 D in LDA and 27 D in TB. The general results in LDA and in TB are quite similar, and agree with recent calculations \cite{Brady99}. Fig. \ref{fig:fig1} describes the original Aviram-Ratner mechanism \cite{Aviram74} for a D-$\sigma$-A molecule. The electronic states are supposed to be totally localized either on the D side or on the A side. The HOMO(D) and LUMO(D) are high in energy, compared to, respectively, the HOMO(A) and LUMO(A). Therefore, a current can be established at relatively small positive bias, such that the Fermi level at the A side is higher than LUMO(A), and the Fermi level at the D side is lower than HOMO(D), providing that the electrons can tunnel inelastically through the $\sigma$ bridge. Thus we expect an asymmetric $I(V)$ curve with an important rise of the current at positive bias above a given threshold (the electrode at A side is grounded, as in the experiments \cite{Metzger97,Vuillaume99,Metzger01}). In the case of our D(+)-$\pi$-A(-) molecules ({\bf 1} and {\bf 2}), LDA and TB calculations show that the HOMO and the LUMO are strongly delocalized on the whole molecule. Even if the HOMO is more localized on the A part and the LUMO on the D part, the extension of the wave functions on the opposite side exceeds 30\%. Thus the two parts of the molecules cannot be decoupled, as in the original Aviram-Ratner model. This is obviously due to the $\pi$ bridge between the donor and acceptor sites. Because of the large dipole of the free molecule (27 D in TB), electrostatic interactions play an important role in the LB film. We plot in Fig. \ref{fig:fig2} the intensity of the dipole, when we vary the molecular area in a free monolayer without a metallic electrode. At a molecular area of 58.5 \AA$^{2}$, the dipole is screened by a factor 2.6, due to the Coulomb interaction between the parallel dipoles. When the metallic electrodes are included, the dipole moment is enhanced, because of the attractive interaction between the charges in the dipole and those induced in the electrodes. We obtain a final dipole of 19 D in the diode at zero bias. This dipole layer gives rise to a built-in potential between the two sides of the film, as shown in Fig. \ref{fig:fig3}, where we plot the average electrostatic potential in the system. In presence of the electrodes and at zero bias, opposite charges appear on  the electrodes (5.8$\times$10$^{-03}$ nelectrons/ \AA$^{2}$), so that the induced potential drop exactly cancels the one created by the dipole layer. When a bias is applied to the diode, the HOMO and the LUMO energies become linear functions of the voltage $(V)$, with the same slope ($\eta$) for the two levels (Fig. \ref{fig:fig4}), which is another consequence of their delocalization. Indeed, if the two states wre localized at opposite sides of the molecule, the energy of the state closer to the grounded electrode would vary less than the other. For example, it was shown recently on D-$\sigma$-A molecules that the HOMO-LUMO gap is strongly distorted by the electric field \cite{Brady99}, because the two states are localized on different sites. In contrast, in our case of a D-$\pi$-A molecule, the gap is almost independent of the applied voltage. The slope $\eta$, obtained from the self-consistent calculation, strongly depends on the length of the alkyl chain (Fig. \ref{fig:fig5}). It can be estimated, by comparing the center of the chromophore of the molecule with the molecular length. Thus, for Q-3CNQ, $\eta = 0.49$ (close to 1/2), because the potential drop is symmetrically shared in the molecule, while for molecule {\bf 1} it is 0.21 (close to 1/4), because the states are localized in the Q-3CNQ part, close to the grounded electrode, whereas a large part of the potential drop takes place in the alkyl chains, which have a small polarizability due to their large gap \cite{Vuillaume98}. This important effect is particularly striking in Fig. \ref{fig:fig3}, where we compare the electrostatic potential in the system at zero bias and at +2 V. The voltage drop in the molecular layer, and thus the factor $\eta$, can be simulated, using a simple dielectric model, where the different parts of the structure are described by a dielectric medium. We consider three regions: the vacuum gaps between the molecules and the metal electrodes, the Q-3CNQ part, and the alkyl chain, with respective widths $d_{vacuum}$, d$_{\pi}$, d$_{alkyl}$, and respective dielectric constants  $\epsilon_{\pi}$ and $\epsilon_{alkyl}$. We obtain easily:

\begin{equation}
\eta=\frac{1}{2}\cdot \frac{1}{1+\displaystyle \frac{\epsilon_{\pi}\cdot d_{alkyl}}{\displaystyle \epsilon_{alkyl \cdot (d_{\pi}+\displaystyle d_{vaccuum}\cdot \epsilon_{\pi})}}}
\end{equation}

This equation explains well the dependence of $\eta$ with the length of the alkyl chain (Fig. \ref{fig:fig5}). The best fit is obtained with $\epsilon_{\pi} = 10.0$ and $\epsilon_{alkyl} = 1.6$, in agreement with the fact that the alkyl chains have a small polarizability. Measurements on monolayers of alkyl chains between metal electrodes give values of $\epsilon_{alkyl}$ between 2.0 and 2.5, but for a density of one molecule per surface of 20-22 \AA$^{2}$  \quad \cite{Ulman91} (compared to 58.5 \AA$^{2}$ in our case). Our calculations for such a monolayer with a molecular area of 20 \AA$^{2}$ gives $\epsilon_{alkyl}= 2.1$ , in agreement with the experiments.

\section{CALCULATION OF THE CURRENT}
The current intensity is calculated using the Landauer formula [18]

\begin{equation}
I=\frac{2e}{h}\int_{-\infty}^{+\infty}[f_{L}(\epsilon)-f_{R}(\epsilon)]Tr \{G^{a}(\epsilon)\Gamma_{R}(\epsilon)G^{r}(\epsilon)\Gamma_{L}(\epsilon)\} d\epsilon
\label{eq:eq3}
\end{equation}

where $f_{L}(\epsilon)$ and $f_{R}(\epsilon)$ are the Fermi-Dirac distribution functions in the left and right leads, respectively.  $G^{r,a}(\epsilon)$ are the retarded and advanced Green's functions of the molecule, and the $\Gamma^{L,R}(\epsilon)$ matrices that describe the coupling of the molecule to the electrodes \cite{Meir92}. The Green's functions of the semi-infinite metal electrodes are calculated in TB \cite{note1}, using the decimation method \cite{Guinea83}. The hopping matrix elements between metal orbitals ($\gamma$) and molecular orbitals ($\beta$) are restricted to the interactions with metal atoms of the top surfaces. They are written

\begin{equation}
V_{\gamma \beta}=\frac{S_{\gamma \beta}(d)}{S_{\gamma \beta}(d_{0})}V^{H}_{\gamma \beta}(d_{0})
\label{eq:eq4}
\end{equation}

where $d$ is the inter-atomic distance, $d_{0}$ is the sum of the covalent radii of the two atoms, and $V^{H}_{\gamma \beta}(d_{0})$ is the hopping integral, calculated from Harrison's rules \cite{Harrison80} at the inter-atomic distance is the hopping matrix element, as calculated by Chen \cite{Chen93} to study the current in a scanning tunneling microscope. The advantages of expression (\ref{eq:eq4}) are: i) at small distance $d$ close to $d_{0}$, the couplings are close to Harrison's values, which give a good description of the chemical interactions \cite{Harrison80}; ii) at larger distances $(d >> d_{0})$, the exponential dependence of the hopping integrals with $d$ is correctly described by the expressions given by Chen \cite{Chen93}. For a given bias, we calculate the self-consistent electronic structure and the Green's functions of the molecule in the LB film. The Green's functions of the whole system are calculated by coupling the molecules to the leads.

\section{Results for Al Electrodes}
Let us consider the case of Al electrodes. We need to discuss the problem of the position of the molecular levels with respect to the Fermi level of the electrodes at zero bias. We have considered three situations (a, b, c), which are summarized in Fig. \ref{fig:fig6}. The first situation (a) corresponds to the vacuum level alignment, a common approximation for physisorbed molecules. The HOMO and LUMO energy levels are fixed, with respect to the metal work function (4.2 eV), by the ionization potential $I$ and the electron affinity $A$, respectively. $A$ and $I$ include the charging energy of the molecule, corresponding to its ionization when the current flows through the structure. We have calculated these values in LDA ($I$ = 6.9 eV, $A$ = 2.7 eV for {\bf 1} and {\bf 2}). But, for an ionized molecule in the diode, we must also consider the screening due to the electrodes, i.e. the interaction energy between the extra charge (an electron in the LUMO $\psi_{e}$ or a hole in the HOMO   $\psi_{h}$  and the image charges induced in the electrodes \cite{note2}. It is given fairly accurately, in first-order perturbation theory, by  $\Sigma_{h(e)}=<\psi_{h(e)}|V_{ind}|\psi_{h(e)}>$/2 , where V$_{ind}$ is the electrostatic energy of a charge between two metallic electrodes. We obtain in TB $\Sigma_{e} = 0.53 eV$ and $\Sigma_{h} = 0.62 eV$ for {\bf 1} (resp. 0.90 eV and 0.97 eV for {\bf 2}). We fix the molecular levels by applying a rigid shift to all unoccupied (occupied) states, such that the LUMO (HOMO) level position is given by $A$ + $\Sigma_{e}$ ($I$- $\Sigma_{h}$ ). Fig. \ref{fig:fig6}  (case a) shows that the Al Fermi level is closer to the LUMO level than the HOMO. In fact, the vacuum level alignment does not occur in many metal-organic systems \cite{Ishii99}. In our case, the alignment may be influenced by charge distributions in the thin Al oxide, which is unavoidably present between the Al electrodes and the LB monolayer \cite{Vuillaume99}. The main effect is a rigid shift of the whole set of molecular energy levels \cite{Ishii99}. Thus, in order to investigate a broad range of possible experimental situations, we have considered a second situation (b) corresponding to the Al Fermi level at the middle of the HOMO-LUMO gap and, a third situation (c), symmetric to the first one, where the Al Fermi level is closer to the LUMO level. The calculated $I(V)$ characteristics are plotted in Fig. \ref{fig:fig7} for molecule {\bf 1}, with its long alkyl chain. In the level configuration (a), we predict an asymmetric $I(V)$ curve with a forward behavior at negative bias, contrary to the Aviram-Ratner situation. The onset of the current at -1.1 V corresponds to the resonance of the LUMO level with the Fermi level of the D(+)-side electrode, as shown in Fig. \ref{fig:fig4}. The electrical rectification arises from the asymmetric profile of the electrostatic potential across the system, leading to $\eta \ne \frac{1}{2}$. The resonance of the LUMO with the Fermi level of the A(-)-side electrode could only occur at high positive bias (3.7 V). In configuration (b), the resonance with the LUMO shifts to a more negative voltage (-1.8 V) but there is an additional onset at +1.8 V, corresponding to the resonance with the HOMO level (almost invisible in Fig. \ref{fig:fig7} because the transmission coefficient is small). Finally, in configuration (c), only this latter resonance is possible at +1.0 V, giving rise to a rectification at positive bias. The situation is completely different in the case of Q-3CNQ (inset of Fig. \ref{fig:fig7}), where $\eta \approx  \frac{1}{2}$. If a Fermi level of an electrode energetically reaches some molecular level at a given bias, the same effect occurs symmetrically with the Fermi level of the other electrode at the opposite bias (Fig. \ref{fig:fig4}). This leads to -almost \cite{note3}- symmetric I(V) curves {\it for the three configurations (a, b, c)}. Experimental studies \cite{Metzger97,Vuillaume99,Metzger99} have been made on molecules {\bf 1} with a long alkyl chain; some Al$|$monolayer$|$Al devices show (i) a rectification at positive bias and relatively low current \cite{Metzger99},(ii) some have symmetrical $I(V)$ curves, and (iii) some have relatively high currents and rectification at negative bias \cite{Vuillaume99}. The calculations presented here would make case (i) correspond to configuration c, where the HOMO level is close to the metal Fermi level, and case (iii) correspond to configuration a, where the LUMO level crosses the Fermi level of the Al electrode with bias  applied to it. All this experimental variability is either due to molecules turning upside down after film deposition and before measurement \cite{Metzger97}, or it is due to non-uniform charges in the thin Al oxide \cite{Metzger99}.

\section{Results for Au Electrodes}
Recent experiments \cite{Metzger01} have been performed on LB films of the same molecules but using oxide-free Au electrodes. All the operating devices (not short-circuited) are rectifying at positive bias with a threshold at $\sim$ +1.6 V (Fig. \ref{fig:fig8}). The main difference between Au and Al electrodes comes from the metal work function (resp. 5.3 eV and 4.2 eV). Shifting the zero bias Fermi level at -5.3 eV, we can just transpose the results of Fig. \ref{fig:fig4}, from which we calculate the voltages for the resonance between the HOMO-LUMO levels and the Fermi levels. As oxide charges cannot be invoked in the case of Au electrodes, we have only considered the situation (a) where levels are aligned with respect to the vacuum level. We obtain in this case a rectification at positive bias (+1.24 V) in agreement with the experiments, corresponding to the resonance of the HOMO with the Fermi level at the D(+)-side electrode (situation close to the case c for Al). The resonance with the Fermi level at the A(-)-side only occurs at -4.6 V. We compare in Fig. \ref{fig:fig8} one typical experimental $I(V)$ curve \cite{Metzger01} with the calculated one for Au$|$C$_{16}$H$_{33}$Q-3CNQ$|$Au. As discussed above, the theory predicts a rectification at positive bias, {\it in agreement with the experiments}. But we have been obliged to multiply the calculated current by a factor 10$^{3}$ to be on the experimental scale. We have checked that there is no way to explain this factor by reducing the distance between the molecules and the electrodes. Thus one possible explanation is that some metal of the top Au electrode deposited on the LB film has slightly diffused in the molecular layer. This is equivalent to a reduction of the length of the alkyl chain, which acts as a barrier for the tunneling of the electrons. Fig. \ref{fig:fig8} shows that a correct magnitude of the current is obtained for Au$|$C$_{8}$H$_{17}$Q-3CNQ$|$Au, or equivalently for a diffusion of gold over a length of $\sim$ 4 \AA (taking into account the tilting of the molecules). This is quite realistic with gold electrodes, even if the experimental procedure of Ref. \cite{Metzger01} has been designed to avoid at maximum the diffusion of the metal. The theory predicts a saturation of the current at high voltage, which is not observed experimentally as the experimental threshold ($\sim$ +1.6 V) is at higher voltage. We must add that the step-like dependence of the calculated current would be broadened by taking into account the interactions between the molecules or the coupling to molecular vibrations. On the experimental side, the application of high bias starts to degrade the molecular diode  \cite{Metzger01}.

\section{Conclusion}
In conclusion, we have presented detailed calculations of the $I(V)$ characteristics in a LB film of C$_{16}$H$_{33}$Q-3CNQ sandwiched between two Al or Au electrodes. We show the importance to include Coulomb interactions and screening effects in a self-consistent manner. Inescapable conclusions are that the placement of the electroactive part of the molecule within the gap between metal electrodes is very important, and that an important experimental issue in the future will be to control the band alignment at the organic-metal interface.

\begin{acknowledgments}
The Institut d'Electronique et de Micro\'electronique du Nord is UMR 8520 of CNRS.

\end{acknowledgments}

\clearpage

\noindent{\bf FIGURES}

\begin{figure}
\caption{\label{fig:fig1}  Top (a): Schematic representation of C$_{16}$H$_{33}$Q-3CNQ and of the diode. Bottom (b): the Aviram-Ratner mechanism for molecular rectification.}
\end{figure}

\begin{figure}
\caption{\label{fig:fig2} Evolution of the dipole moment of C$_{16}$H$_{33}$Q-3CNQ in a monolayer without electrode as a function of the molecular area.}
\end{figure}

\begin{figure}
\caption{\label{fig:fig3} Electrostatic potential in the metal $|$ C$_{16}$H$_{33}$Q-3CNQ film $|$ metal system at zero bias (dashed line) and at +2 V (straight line). The potential is defined as an average value in a lattice unit cell.}
\end{figure}

\begin{figure}
\caption{ \label{fig:fig4} Evolution of the LUMO and HOMO energies (relative to the vacuum level) as a function of the applied bias (straight lines: C$_{16}$H$_{33}$Q-3CNQ, dotted lines: Q-3CNQ). At zero bias, the molecular levels and the metal Fermi levels are as shown for case a of Fig. \ref{fig:fig6}. The crossings of the LUMO and HOMO energies with the Fermi levels (dashed lines) correspond to current thresholds in the $I(V)$ characteristics.}
\end{figure}

\begin{figure}
\caption{\label{fig:fig5} Variation of $\eta$ with the length of the molecules C$_{x}$H$_{2x+1}$Q-3CNQ (* full self-consistent tight binding calculation; straight line: simple dielectric model). $\eta$ is the slope of the linear dependence of the HOMO and the LUMO energies with respect to the applied voltage.}
\end{figure}

\begin{figure}
\caption{\label{fig:fig6} Three possible schemes for the position of the molecular energy levels with respect to the Fermi level of the leads at zero bias. I and A are, respectively, the ionization energy and the electron affinity of the isolated molecules. $\Sigma_{h}$ and $\Sigma_{e}$ account for the reduction of the charging energy in the diode due to the screening by the electrodes.}
\end{figure}

\begin{figure}
\caption{ \label{fig:fig7} Calculated I(V) characteristics of Al$|$C$_{16}$H$_{33}$Q-3CNQ$|$Al for the three configurations of the molecular levels, with respect to the Fermi level of the electrodes, as defined in Fig. 6 (straight line: a, dashed line: b, dotted line: c). Insets: same for Al$|$Q-3CNQ$|$Al.}
\end{figure}

\begin{figure}
\caption{ \label{fig:fig8} Experimental \cite{Metzger01} and calculated current-voltage curves of molecular diodes with Au electrodes.}
\end{figure}

\cleardoublepage

\newpage
\includegraphics[scale=0.8]{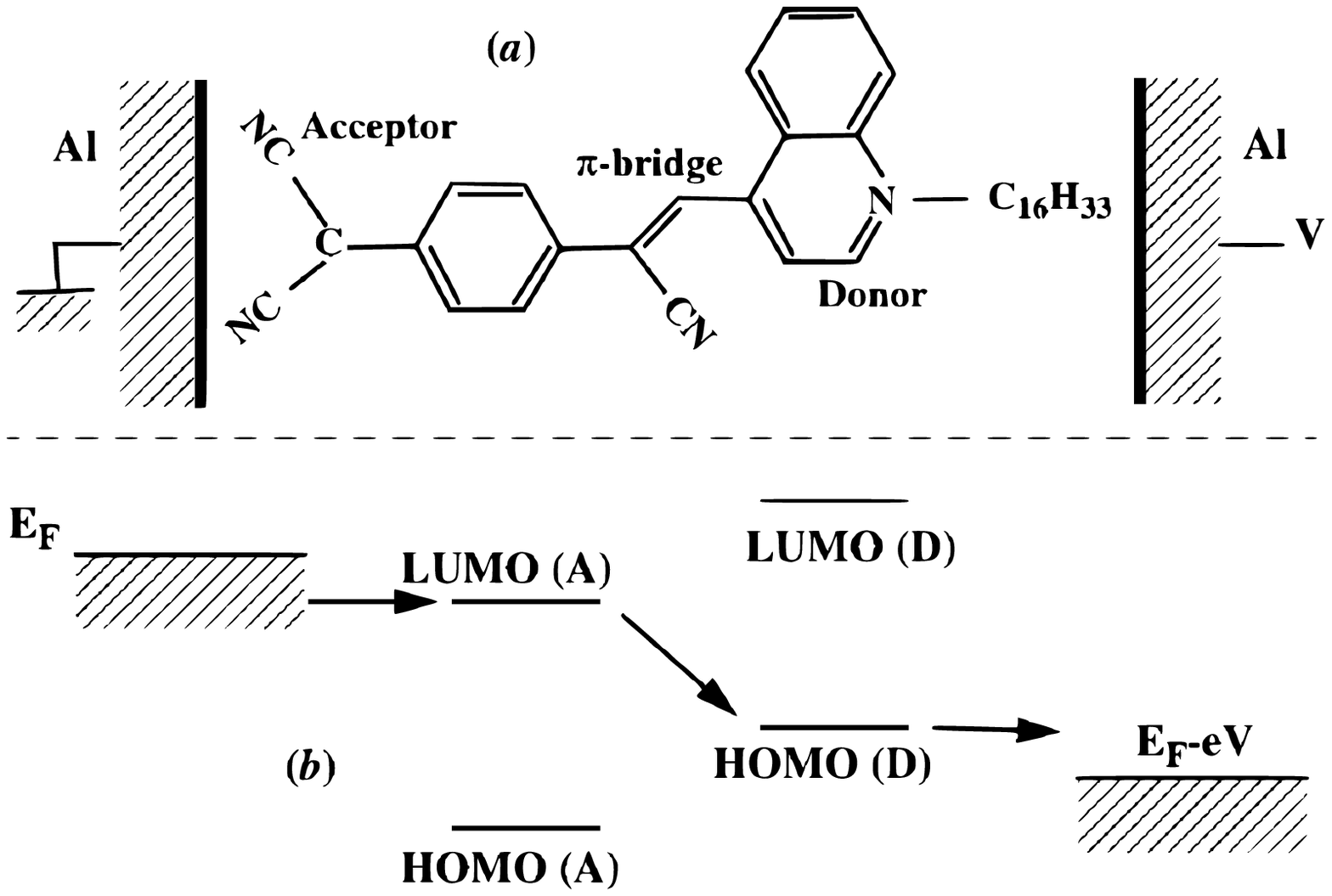}
\newpage
\includegraphics[scale=0.8]{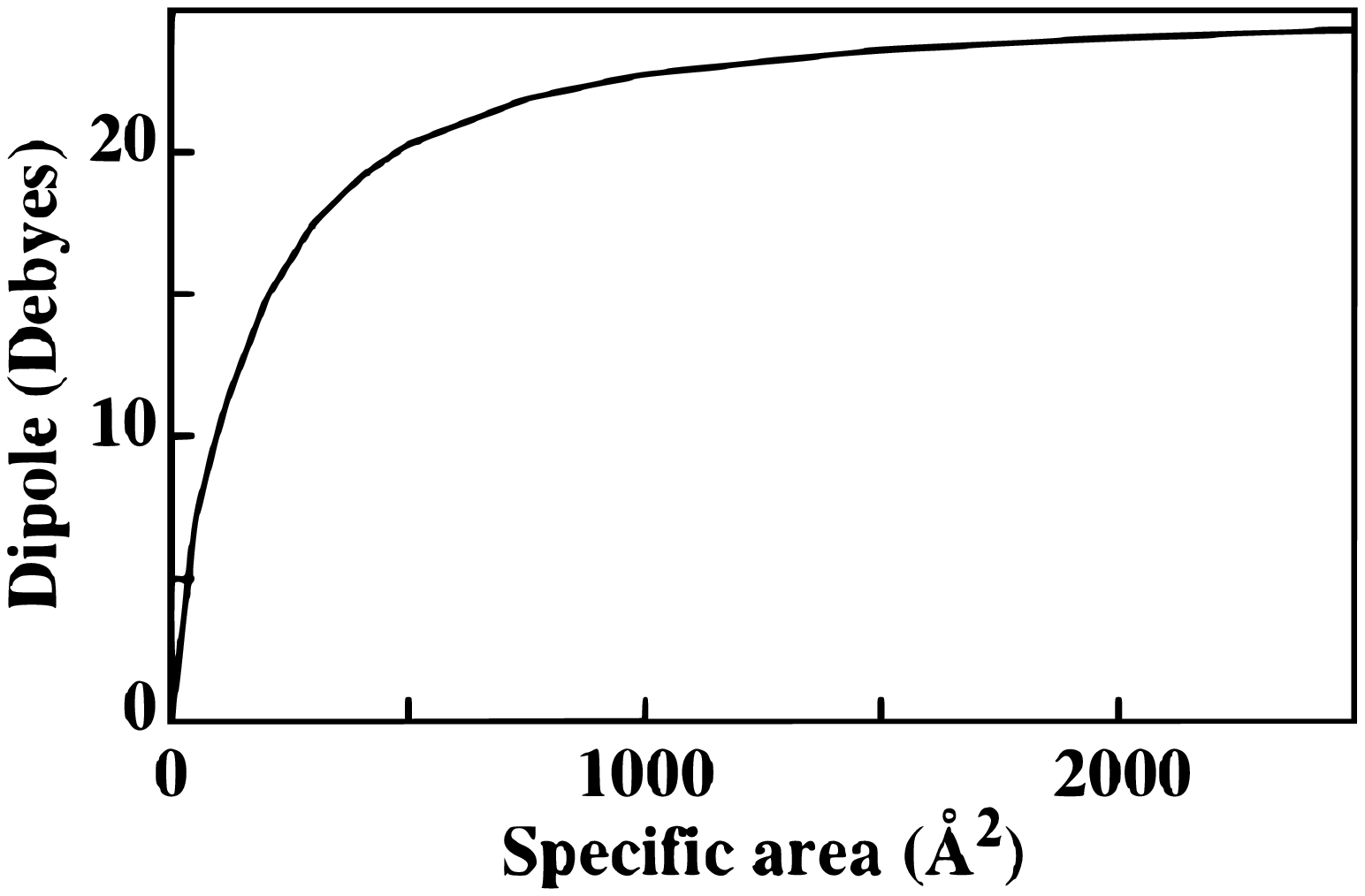}
\newpage
\includegraphics[scale=1.0]{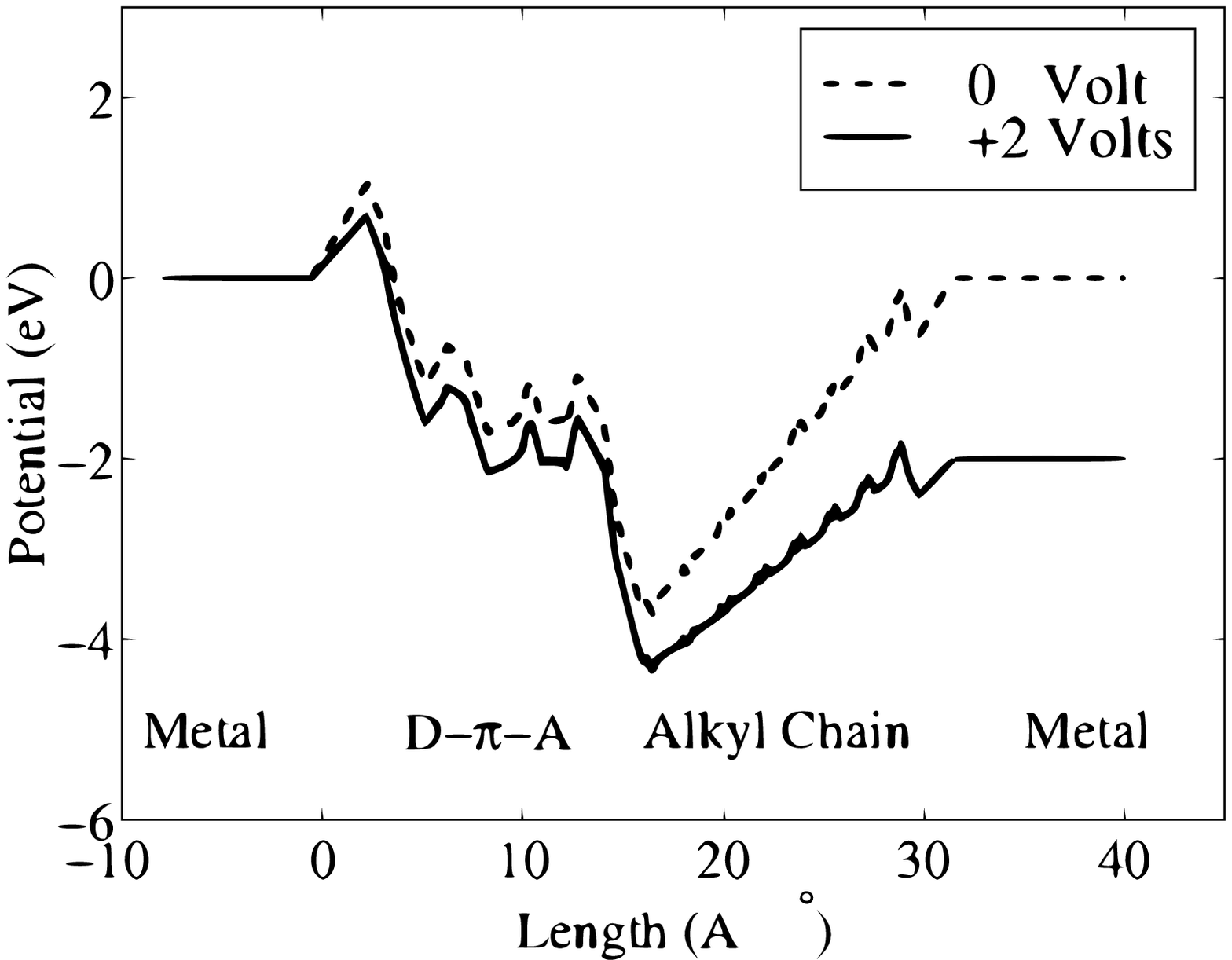}
\newpage
\includegraphics[scale=0.8]{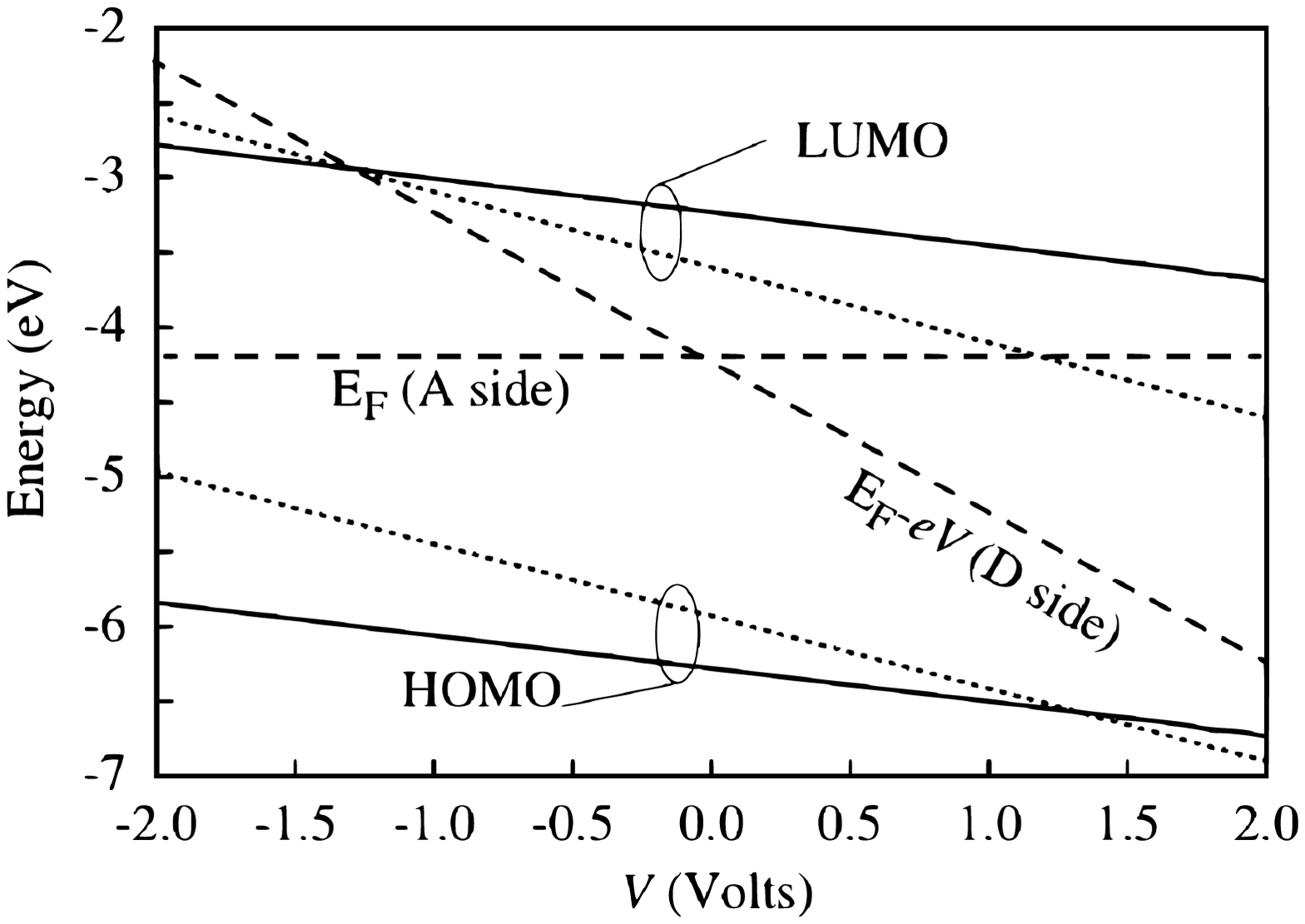}
\newpage
\includegraphics[scale=1.0]{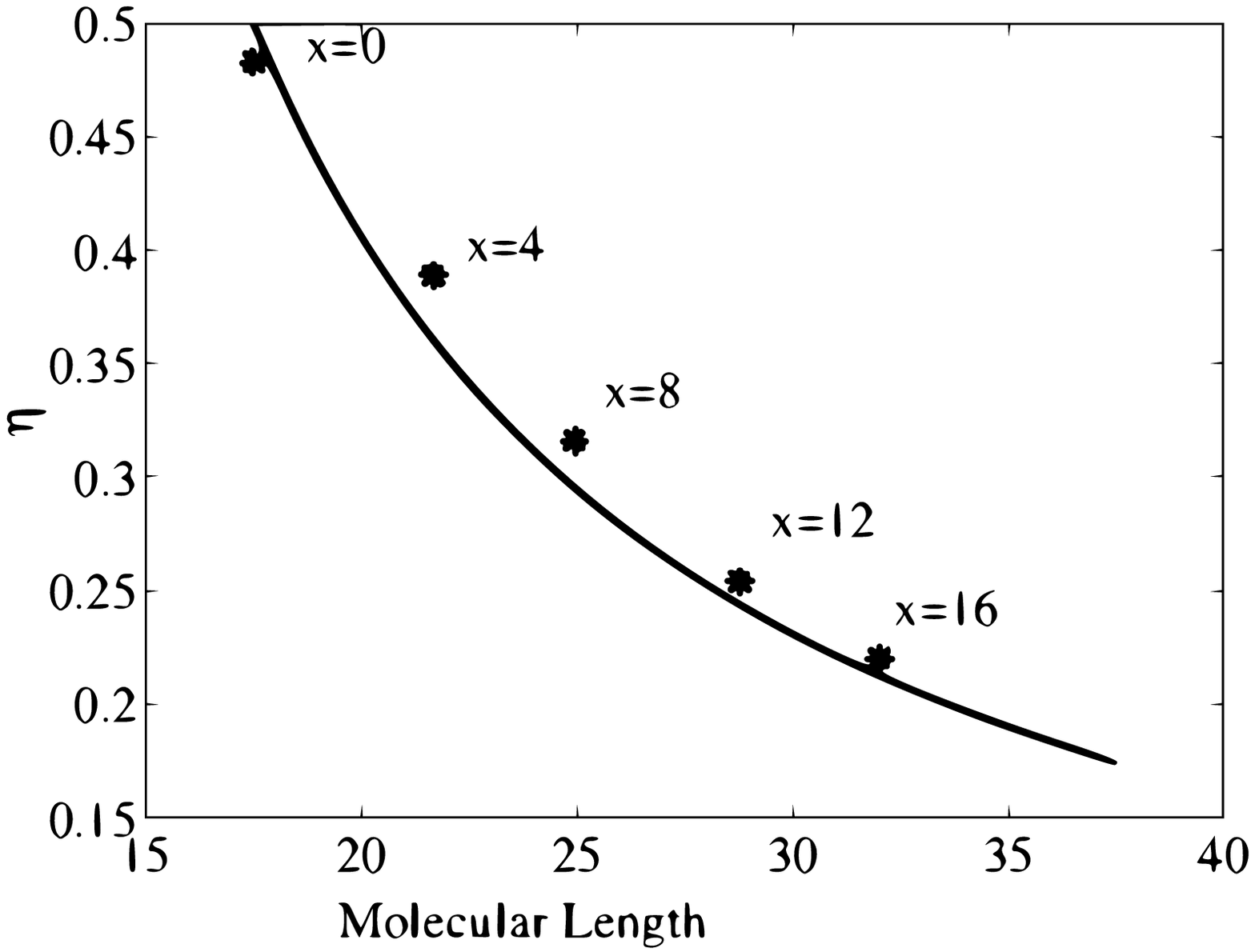}
\newpage
\includegraphics[scale=1.0]{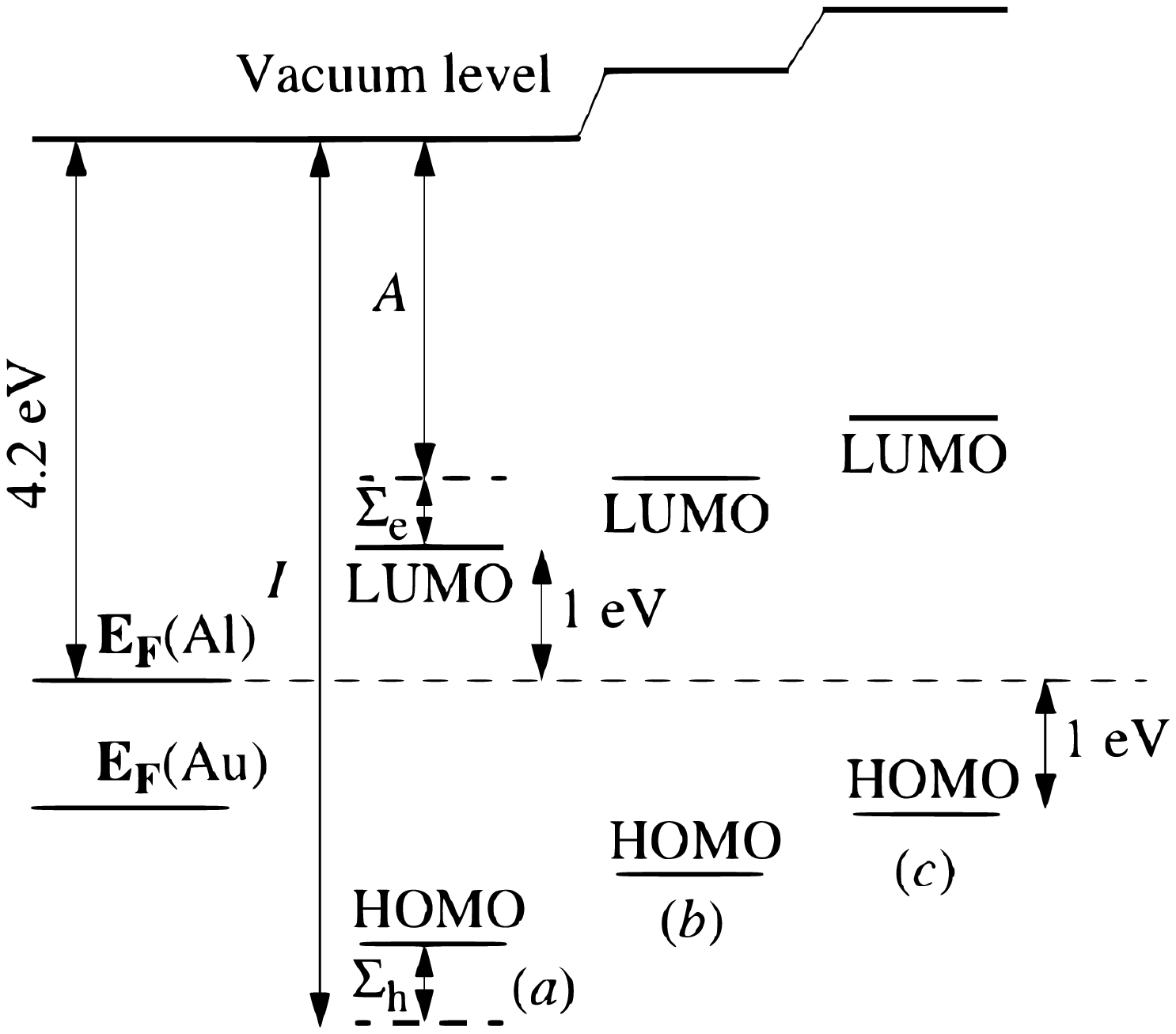}
\newpage
\includegraphics[scale=1.0]{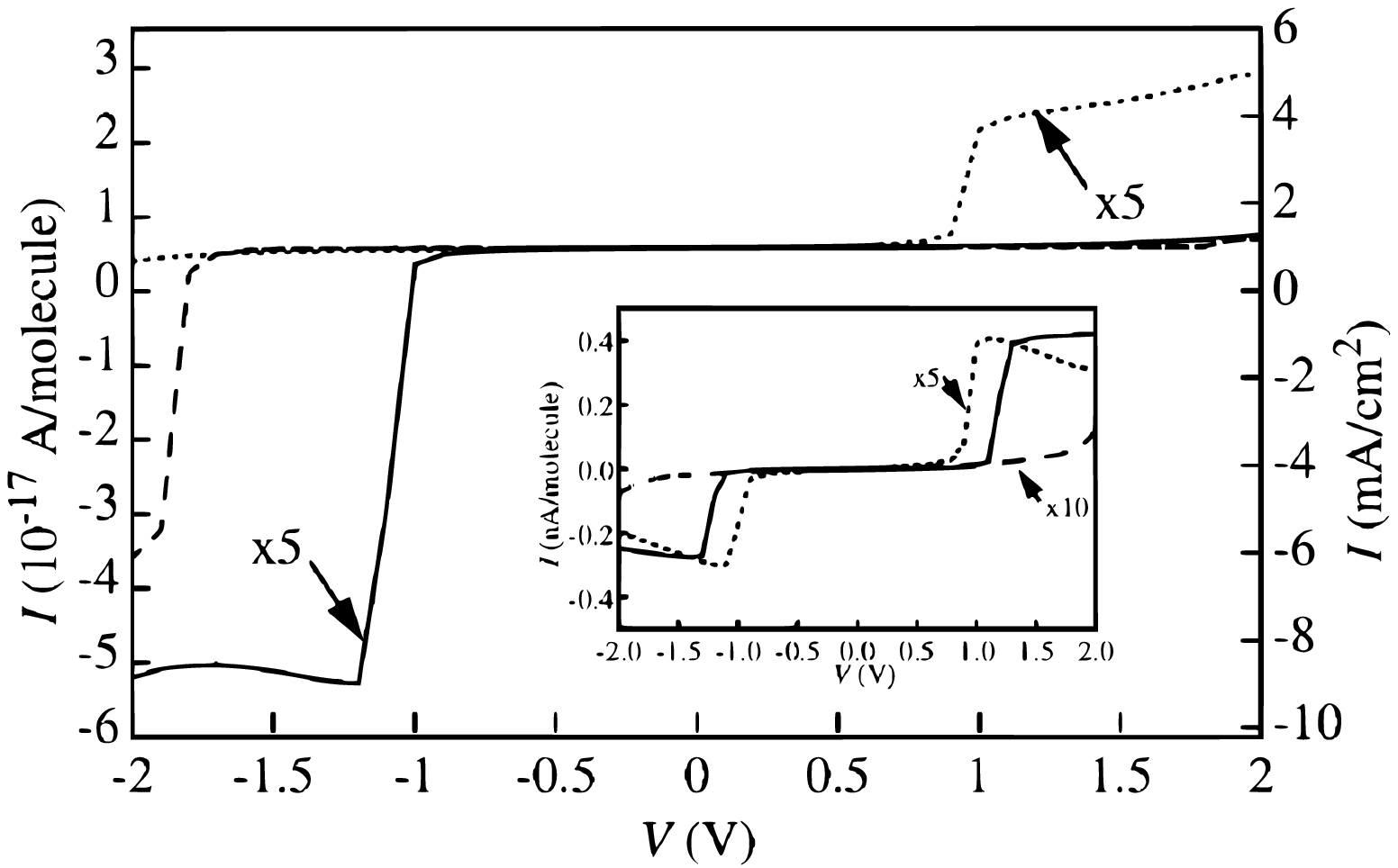}
\newpage
\includegraphics[scale=1.0]{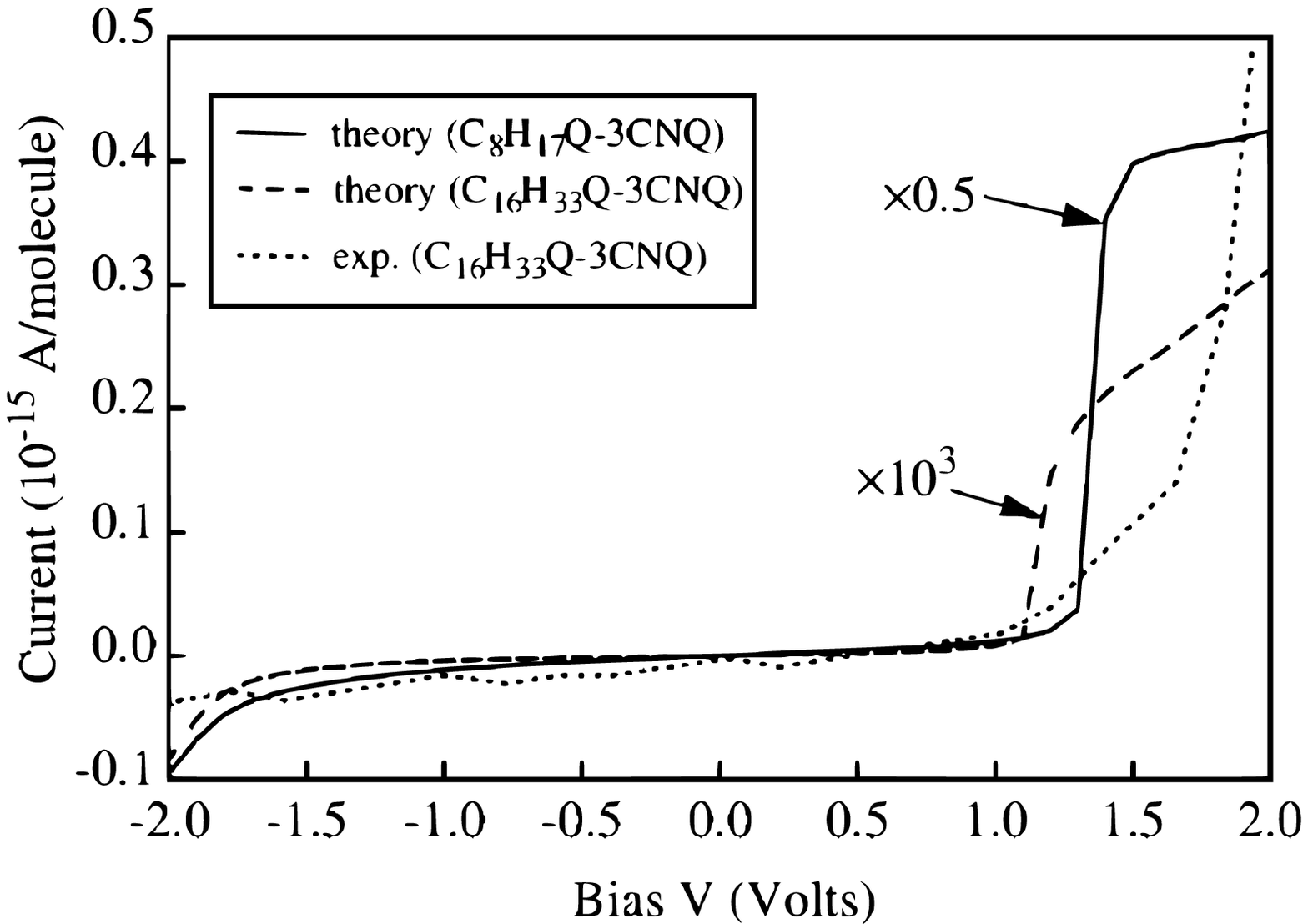}


\begin{thebibliography}{0}
\expandafter\ifx\csname natexlab\endcsname\relax\def\natexlab#1{#1}\fi
\expandafter\ifx\csname bibnamefont\endcsname\relax
  \def\bibnamefont#1{#1}\fi
\expandafter\ifx\csname bibfnamefont\endcsname\relax
  \def\bibfnamefont#1{#1}\fi
\expandafter\ifx\csname citenamefont\endcsname\relax
  \def\citenamefont#1{#1}\fi
\expandafter\ifx\csname url\endcsname\relax
  \def\url#1{\texttt{#1}}\fi
\expandafter\ifx\csname urlprefix\endcsname\relax\def\urlprefix{URL }\fi
\providecommand{\bibinfo}[2]{#2}
\providecommand{\eprint}[2][]{\url{#2}}

\end{thebibliography}


\newpage

\begin{thebibliography}{}

\bibitem{Aviram74} A. Aviram and M. Ratner, Chem. Phys. Lett. 29, 277 (1974).

\bibitem{Ashwell93} G. J. Ashwell, J. R. Sambles, A. S. Martin, W. G. Parker, and M. Szablewski, J. Chem. Soc., Chem. Commun. 1374 (1990); A. S. Martin, J. R. Sambles, G. J. Ashwell, Phys. Rev. Lett. 70,
218 (1993).
\bibitem{Metzger97} R. M. Metzger, B. Chen, U. Höpfner, M. V. Lakshmikantham, D. Vuillaume, T. Kawai, X. Wu, H. Tachibana, T.V. Hughes, H. Sakurai, J.W. Baldwin, C. Hosh, M.P. Cava, L. Brehmer, and
C.J. Ashwell, J. Am. Chem. Soc. 119, 10455 (1997).
\bibitem{Vuillaume99} D. Vuillaume, B. Chen and R.M. Metzger, Langmuir 15, 4011 (1999). B. Chen and R.M.
Metzger, J. Phys. Chem. B 103, 4447 (1999).
\bibitem{Metzger99} R. M. Metzger, J. Mater. Chem. 9, 2027 (1999); {\it ibid} 10, 55 (2000).
\bibitem{Metzger01} R. M. Metzger, T. Xu, and I.R. Peterson, Angewandte Chemie International Ed. Engl., in press.
\bibitem{Cerius2}  Cerius$^{2}$ User Guide. San Diego: Molecular Simulations Inc. (1997).
\bibitem{Vosko80}  S. J. Vosko, L. Wilk, and M. Nusair, Can. J. Phys. 58, 1200 (1980).
\bibitem{Perdew92} J.P Perdew, Y.Wang, Phys. Rev. B 45, 13244 (1992).
\bibitem{Krzeminski99} C. Krzeminski, C. Delerue, G. Allan, V. Haghet, D. Sti\'evenard, E. Levillain, and J. Roncali, J. Chem. Phys. 111, 6643 (1999).
\bibitem{Lannoo74} M. Lannoo, Phys. Rev. B 10, 2544 (1974).
\bibitem{Ewald21} P.P. Ewald, Ann. Phys. (Leipzig) 54, 253 (1921).
\bibitem{Koch99}  N. Koch, L.-M. Yu, V. Parente, R. Lazzaroni, R.L. Johnson, G. Leiring, J.-J. Pireaux, J.-L. Br\'edas, Synthetic Metals 101, 438 (1999).
\bibitem{Kwon99} O. Kwon, M.L. McKee, and R. Metzger, Chem. Phys. Letters 313, 321 (1999).
\bibitem{Brady99} A.C. Brady, B. Hodder, A.S. Martin, J. R. Sambles, C.P. Ewels, R. Jones, P. R. Briddon, A.M.
Musa, C.A. Panetta, and D.L. Mattern, J. Mater. Chem. 9, 2271 (1999).
\bibitem{Vuillaume98} D. Vuillaume, C. Boulas, J. Collet, G. Allan and C. Delerue, Phys. Rev. B 58, 16491 (1998).
\bibitem{Ulman91} A. Ulman, {\it An introduction to Ultrathin Organic Films: From Langmuir-Blodgett to Self-
Assembly} (Academic, San Diego, 1991).
\bibitem{Meir92} Y. Meir and N.S. Wingreen, Phys. Rev. Lett. 68, 2512 (1992).
\bibitem{note1} Tight binding parameters for Al and Au have been fitted on a LDA band structure.
\bibitem{Guinea83} F.Guinea, J Sanchez-Dehesea and F Flores, J.Phys.C: Solid State Phys.16, 6499 (1983).
\bibitem{Harrison80} W.A. Harrison, {\it Electronic Structure and the Properties of Solids} (Freeman, San Francisco,
1980).
\bibitem{Chen93}  C.J. Chen, {\it Introduction to scanning tunneling microscopy} (Oxford University Press, New
York, 1993).
\bibitem{Yaliraki99} S.N. Yaliraki, M. Kemp, and M.A. Ratner, J. Am. Chem. Soc. 121, 3428 (1999).
\bibitem{note2} Additional screening comes from the polarization of the surrounding molecules. A calculation of the transverse polarizability of the molecules in LDA shows that this effect is small.
\bibitem{Ishii99} H. Ishii, K. Sugiyama, E. Ito, and K. Seki, Adv. Mater. 11, 972 (1999).
\bibitem{note3} The small asymmetry comes from the fact that the right and left sides of the molecule couples
differently to the Al electrodes

\end{thebibliography}
\end{document}